\documentclass[12pt]{article}

\usepackage{amstext,amsmath,amssymb,amsfonts,bbm}
\usepackage[latin1]{inputenc}
\usepackage{epsfig}
\usepackage{hyperref}
\usepackage{amsthm}
\usepackage{subfigure}
\usepackage{color}
\usepackage{multirow}
\newcommand{\bea}{\begin{eqnarray}}
\newcommand{\eea}{\end{eqnarray}}

\newcommand{\cG}{{\mathcal G}}

\begin{document}

 \title{EPRL/FK Group Field Theory}
\author{Joseph Ben Geloun\footnote{The Perimeter Institute for Theoretical Physics, Waterloo, ON, N2L 2Y5, Canada;
International Chair in Mathematical Physics and Applications, ICMPA-UNESCO Chair, 072BP50, Cotonou, Rep. Benin.}, 
Razvan Gurau\footnote{The Perimeter Institute for Theoretical Physics, Waterloo, ON, N2L 2Y5, Canada. }, 
Vincent Rivasseau\footnote{Laboratoire de Physique Th\'eorique, CNRS UMR 8627,
Universit\'e Paris XI,  F-91405 Orsay Cedex, France.}
}

\maketitle
\begin{abstract}
The purpose of this short note is to clarify the Group Field Theory vertex and propagators
corresponding to the EPRL/FK spin foam models and to detail the subtraction of leading divergences 
of the model.
\end{abstract}

\begin{flushright}
pi-qg-194\\
ICMPA-MPA/2010/15
\end{flushright}

\noindent  Pacs: 11.10.Ef, 11.10.Gh\\
\noindent  Key words: group field theory, perturbative expansion, quantum gravity

\medskip

\section{Introduction}

Group field theories (GFTs) (see \cite{boul,sasa1, Freidel} and \cite{oriti, Oriti:2009wn})
are the higher dimensional generalization of random matrix models. Like in matrix models, the Feynman graphs of group 
field theory are dual to triangulations (gluing of simplices). 
The combinatorics of a Feynman graph encodes the topology of the gluing while its amplitude 
encodes a sum over metrics compatible with a fixed gluing. The correlation functions of GFT's sum over both metrics and topologies. 
Generically GFT's generate topological singularities \cite{Gurau:2010nd},
but the most dangerous can be eliminated by restricting the  allowed gluings 
by a ``coloring'' prescription \cite{Gurau:2009tw}.

The metrics appear in GFT through their holonomies, group elements of $SO(D)$
for $D$ dimensional manifolds. The Feynman amplitudes are therefore 
integrals on $SO(D)$, or sums over spin indices in Fourier space. Such amplitudes are also 
referred to as spin foams \cite{rovel}.
Due to the Wick theorem, GFT provide a prescription of the 
class of graphs that should be summed, together with their combinatoric weights.

In any quantum field theory there is some ambiguity in the definition of propagators 
and vertices. A vertex can be dressed by an arbitrary fraction of the propagator 
without changing the bulk theory, provided we amputate each propagator by the square 
of that fraction. What fixes this ambiguity in ordinary quantum field theory 
is a locality requirement on the vertices.

In \cite{KMRTV}, such a locality requirement was proposed for GFTs, namely to restrict 
their vertices to simple products of $\delta$ functions which identify group 
elements in strands crossing the vertex. Everything else should be considered part of 
the propagator. Beware that this is {\it not} the usual spin-foam terminology.
However, as we will see in the sequel, it immediately leads to a well defined
and simple prescription to identify divergences.

In dimension $D$ the simplest and most natural vertex with this locality property
represents $D+1$ subsimplices of dimension $D-1$ bounding a $D$ dimensional 
simplex (hence connected through $D(D+1)/2$ such $\delta$ functions). The fields are functions on 
$SO(D)^D$, and the $D$-stranded propagators represent the gluing of $D$ dimensional
simplices along $D-1$ dimensional subsimplices. 

Using as propagator a diagonal $SO(D)$ gauge averaging projection $T$ (ensuring 
flatness of the  holonomies), the amplitude of a Feynman graph equals the 
partition function of a BF theory discretized on the dual gluing of simplices.
Recently such models have received increased attention and various partial power counting 
results have been established, either for generic three dimensional models 
\cite{Freidel:2009hd,Magnen:2009at} or for colored and linearized models 
\cite{Geloun:2009pe,Geloun:2010nw, Bonzom:2010ar}

Gravity can be seen as a constrained version of BF theory. In line with this approach, new
spin foam rules have been proposed to implement the so called Plebanski simplicity
constrains and reproduce the partition function of fully fledged 4D gravity 
\cite{EPR,LS,FreKra,ELPR}. These new models (referred to as EPRL/FK in this paper)  
mix the left and right part of $SO(4)\simeq SU(2)\times SU(2)$ in a novel way 
and give a central r\^ole to the Immirzi parameter. Amplitudes of particular 
spin foams in the EPRL/FK models, revealing improved UV behavior, have been derived in 
\cite{Perini:2008pd} and recovered in \cite{KMRTV}. 

But, as spin foams are only Feynman graphs of the GFT, one still needs to identify an 
appropriate GFT propagator which generates the EPRL/FK spin foam amplitudes. 
A first step in this direction has been performed in \cite{KMRTV}, were the propagator 
(written in terms of coherent states) was computed as a product
of gauge (T) and simplicity (S) projection operators, $C=TST$\footnote{Beware that different
letters are used in \cite{KMRTV}.}. Note that $C$ has a non trivial spectrum, 
hence is suited for a RG analysis. Some steps have already been performed in 
\cite{KMRTV} to write the EPRL/FK action in terms of group elements,
here we propose another equivalent formulation, free of explicit sums over coherent states,
and which might lead to a transparent saddle point analysis 
for estimating graph amplitudes.

In this paper we obtain the EPRL/FK propagator in group space and
consequently more compact formulas for both the propagator and the Feynman amplitudes 
of the GFT underlying EPRL/FK spin foams. Our formulas are well defined for irrational 
values of the Immirzi parameter and constitute a better starting point for slicing 
the propagator according to its spectrum  (and subsequently a fully 
fledged RG analysis). Although such an analysis is in progress \cite{BGR}, 
in this paper we limit ourselves to a non technical introduction of
this GFT written exclusively in the group variables. 
It must be mentioned that, in a somewhat different perspective, improved GFT's  
\cite{Baratin:2010wi,Oriti:2009wg} have already been proposed to implement 
directly the simplicity constrains. A direct comparison of the
action functionals proposed in \cite{Baratin:2010wi,Oriti:2009wg} and our 
results shows that they are in fact quite different and it is still an open question 
which one (if any) of these GFT's is the best suited to describe gravity.

This paper is organized as follows: section \ref{sec:simpl} details the simplicity 
projector $S$ in direct space in terms of characters and section \ref{sec:prop} presents 
the EPRL/FK propagator. Section \ref{sec:ampl} computes the Feynman amplitudes of arbitrary graphs,
and section \ref{sec:subtr} explains the subtraction of leading divergences. As an added bonus
we confirm the power counting estimate found in \cite{Perini:2008pd, KMRTV} for a particular 
graph very effectively in our new representation. Technical details are presented in two appendices.

\section{The Simplicity Projector $S$}\label{sec:simpl}

The coherent spin states \cite{perelomov} form an over complete basis in the $SU(2)$ representation 
spaces.  The decomposition of an operator over an over complete basis is not unique, thus one has many 
possible choices for the kernel of the EPRL/FK simplicity projector $S$.
In \cite{FreKra} (and subsequently in \cite{KMRTV}), it is taken to be
\bea\label{eq:KTMRV}
S = \sum_{j^+,j^-}  \delta^\gamma_{j}\; d_{j^+ + j^-}
\int dn \;
\vert j^{+}, n\rangle \otimes \vert j^{-}, n\rangle
\langle  j^{+}, n \vert \otimes \langle j^{-}, n\vert \; ,
\eea
with
\bea
 && d_{j^+ + j^-}= 2(j^+ + j^-)+1 \; ,\crcr
 && \delta^\gamma_{j} = \delta_{\frac{j^+}{j^-}  = \frac{1+\gamma}{|1- \gamma|}} =
\delta_{|1- \gamma|j^+= (1+\gamma)j^- } \; .
\eea
This is a perfectly valid choice but it has one major drawback. Although, as $S$ is a projector,
$S^2=S$, the square of eq. (\ref{eq:KTMRV}) is 
\bea\label{eq:S}
 && S=S^2 = \sum_{j^+,j^-}  \delta^\gamma_{j} \; d^2_{j^+ + j^-} \int dn  dn' \crcr
&& 
\vert j^{+}, n\rangle \otimes \vert j^{-}, n\rangle
\langle  j^{+}+j^-, n \vert j^++ j^{-}, n'\rangle
\langle  j^{+}, n' \vert \otimes \langle j^{-}, n'\vert \; ,
\eea
which looks quite different. This discrepancy is explained by the over completeness of the
coherent states basis\footnote{In order to conclude that $S$ is a projector, in \cite{KMRTV} one proves that 
$S^3=S^2$, rather than proving $S^2=S$.}. It the sequel, we choose the representation provided in eq. 
(\ref{eq:S}) as it is better suited for explicit computations.

Remark that the $\delta^\gamma_{j} $ does not really make sense (e.g. if $\gamma$ is irrational)
but should be understood in an asymptotic sense as $j_\pm \to \infty$. This will be detailed later, 
and the formulas we will derive for the amplitudes of the theory  ultimately make sense for any $\gamma$.

It is important to realize that the eq. (\ref{eq:S}) is in fact only a shortened (and somewhat confusing) notation. 
The operator $S$ acts on functions defined on $SO(4)$ which decompose in Fourier modes as
\bea
 f(g)=\sum d_j f^j_{pm} D^j_{pm}(g) \; ,
\eea
hence the {\it matrix} elements $D^j_{pm}(g)$ (and not the vectors $\vert jm\rangle $) are the analog of 
the plane waves. Matrix elements of the operator $S$ therefore join a $D^{j_1}_{p_1m_1}(g_1)$ to 
a $D^{j_2}_{p_2m_2}(g_2)$, hence have two groups of indices $j_1,p_1,m_1$ and $j_2,p_2,m_2$. 
To make matters worse, over $SO(4)\simeq SU(2)\times SU(2)$ each of the above six indices is in fact a double index,
$\vec j_1 = ( j^+_1, j^-_1)$, corresponding respectively to each of the two copies of $SU(2)$. In full detail $S$ writes
\bea\label{eq:truek}
 &&S^{\vec j_1,\vec j_2 }_{(\vec{p}_1,\vec{m}_1); (\vec{p}_2,\vec{m}_2)} =
\delta_{\vec j_1 , \vec j_2} \delta^\gamma_{j_1}\;   d^2_{j^+_{1} + j^-_{1} }  \delta_{ \vec p_1 ,  \vec p_2 } \nonumber\\
&&\int dn dn' \; \langle  \vec  j_1 , \vec m_1 \vert \Big{(} \vert j^+_{1},  n \rangle \otimes \vert j^-_{1}, n \rangle\Big{)}
\langle   j^+_{1} + j^-_{1} ,  n \vert     j^+_{1} + j^-_{1}  , n' \rangle \nonumber\\
&&\Big{(}\langle  j^+_{1}, n' \vert \otimes \langle j^-_{1}, n'\vert \Big{)} \vert \vec j_2, \vec m_2  \rangle \;,
\eea
where $ \vert \vec j,\vec m \rangle =  \vert j^+, m^+ \rangle \otimes \vert j^-, m^- \rangle$.
Denoting the matrix elements of unitary representations of $SU(2) \times SU(2)$ as
\bea
 D^{\vec j}_{\vec p\, \vec m}\,(g) := D^{j^+}_{p^+m^+}(g^+) D^{j^-}_{p^-m^-}(g^-) \;,
\eea
we find in the direct (group) space 
\bea
 S(g_1,g_2)= \sum d_{j^+_1} d_{j^-_1} S^{\vec j_1 \vec j_2}_{( \vec p_1 \vec m_1);( \vec p_2 \vec m_2)}
D^{\vec j_1}_{\vec p_1\vec m_1}(g_1) \overline{D^{\vec j_2}_{\vec p_2\vec m_2}(g_2)} \; .
\eea
Substituting eq. (\ref{eq:truek}) yields
\bea \label{Sg}
&&S( g_1,  g_2) = \sum d_{j^+_1} d_{j^-_1}
D^{\vec j_1}_{\vec p_1\vec m_1}(g_1) \overline{D^{\vec j_2}_{\vec p_2\vec m_2}(g_2)} 
\delta_{\vec j_1 , \vec j_2} \delta^\gamma_{j_1}\;   d^2_{j^+_{1} + j^-_{1} }  \delta_{ \vec p_1 ,  \vec p_2 } \nonumber\\
&&\int dn dn' \; \langle  \vec  j_1 , \vec m_1 \vert \Big{(} \vert j^+_{1},  n \rangle \otimes \vert j^-_{1}, n \rangle\Big{)}
\langle   j^+_{1} + j^-_{1} ,  n \vert     j^+_{1} + j^-_{1}  , n' \rangle \nonumber\\
&&\Big{(}\langle  j^+_{1}, n' \vert \otimes \langle j^-_{1}, n'\vert \Big{)} \vert \vec j_2, \vec m_2  \rangle  \; ,
\eea
and summing over $\vec p_2$ and $\vec j_2$ (and renaming $\vec j_1 = \vec j$), we get 
\bea \label{eq:mese}
&& S( g_1,  g_2) = \sum d_{j^+} d_{j^-} \,\delta^\gamma_{j}\;   d^2_{j^+ + j^- } \,
D^{\vec j}_{\vec m_2 \vec m_1} \bigl( (g_2)^{-1} g_1 \bigr) 
 {\mathcal I}(\vec  j , \vec m_1,\vec m_2) \; ,
\eea
with
\bea \label{eq:Idef}
{\mathcal I}(\vec  j , \vec m_1,\vec m_2) &&=
\int dn dk \; \langle  \vec  j , \vec m_1 \vert \Big{(} \vert j^+,  n \rangle \otimes \vert j^-, n \rangle\Big{)}
\crcr&&\langle   j^+ + j^- ,  n \vert     j^+ + j^-  , k \rangle 
\Big{(}\langle  j^+, k \vert \otimes \langle j^-, k\vert \Big{)} \vert \vec j, \vec m_2  \rangle \;.
\eea
The integral ${\mathcal I}(\vec  j , \vec m_1,\vec m_2)$ is evaluated in Appendix \ref{app:B}. 
Substituting eq. (\ref{ijmn2}) yields
\bea\label{eq:Sinter}
 S(g_1,g_2) && =
\sum_{j^+,j^-} d_{j^+} d_{j^-}  d_{j^++j^-}
\,\delta^\gamma_{j} \crcr
&& \sum_{\vec m_1, \vec m_2}  D^{j^+ }_{ m^+_2,m^+_1} \bigl( (g_{2}^+)^{-1} g_1^+ \bigr) \; 
D^{j^- }_{ m^-_2, m^-_1} \bigl( (g_{2}^-)^{-1} g_1^- \bigr) \crcr
&& \sum_{r} \int dh \; D^{j^+}_{m^+_1m^+_2}(h) \; D^{j^-}_{m^-_1m^-_2}(h)\; D^{j^++j^-}_{-r-r}(h) \; ,
\eea
where the integral in the last line is performed over only one group element $h\in SU(2)$. Note that due to the
selection rules the sum over $r$ is in fact restricted to a single term $r=m_1^++m_1^-=m_2^++m_2^-$. However,
although this remark is important in a detailed slice analysis of $S$ (hence of the propagator $C$)
we ignore it throughout this paper. The rationale behind this is that, as will be clear in the sequel, 
allowing this fake sum to survive yields canonical expressions in terms of the familiar $SU(2)$ group 
characters for all relevant quantities.
Summing over $\vec m_1, \vec m_2$ and $r$ in eq. (\ref{eq:Sinter}) yields the compact expression 
\bea\label{eq:Sgroup}
S(g_1,g_2) &=&  \sum d_{j^+} d_{j^-} d_{j^++j^-} \,\delta^\gamma_{j} \crcr
&&\int dh\, \chi^{j^+} \bigl( (g^+_2)^{-1}g^+_{1} h \bigr)\, \chi^{j^-} \bigl( (g^-_2)^{-1}g^-_{1} h \bigr) 
\, \chi^{j^++j^-}(h)\crcr
&=&  \sum d_{j^+} d_{j^-} \; d_{J} \; \delta^\gamma_{j}  \delta_{J = j^++j^-} \crcr
&&\int dh\, \chi^{j^+} \bigl( g^+_{1} h(g^+_2)^{-1} \bigr)\, \chi^{j^-} \bigl(g^-_{1} h(g^-_2)^{-1} \bigr) 
\, \chi^{J}(h) \; ,
\eea
with $\chi^j(g)= \text{Tr}_j(g)=\sum_{k} D^j_{kk}(g)$ the character of $g$ in the representation $j$. Note that at this stage
all the coherent state integrals have been performed, and $S$ is written exclusively in terms of group integrals and
characters. Using $\overline{\chi(h)} = \chi(h^{\dagger})$ and the orthogonality of characters
\bea\label{eq:orthchar}
 \int dp \;\chi^{j}(ap^{-1}) \chi^{j'}(pb)= \frac{1}{d_j} \delta_{jj'} \chi^j(ab) \; ,
\eea
one can check directly using eq. (\ref{eq:Sgroup}) that S is a projector. Note that eq. (\ref{eq:Sgroup}) 
makes sense for any value of $\gamma$ as the character of a group element 
$\chi^j(g)=\frac{\sin(j+\frac{1}{2})\theta}{\sin\frac{\theta}{2}}$ is well defined for all 
values of $j$, half integer or not. 

The simplicity projector $S$ admits several limiting cases
\begin{itemize}
 \item $\gamma=1$ sets $j^-=0$, and $S$ becomes
  \bea
  S(g_1,g_2)&=& \sum_{j^+,J} d_{j^+} d_{J}\delta_{J = j^+}
\int dh\, \chi^{j^+} \bigl(g^+_{1}h (g^+_2)^{-1}\bigr) \chi^{J}(h) \crcr 
&=& \sum_{J} d_J \chi^{J}(g^+_{1} (g^+_2)^{-1})\; = \; \delta(g^+_{1} (g^+_2)^{-1} ) \; ,
  \eea
leading to a BF theory for the $+$ copy of $SU(2)$.
 \item Ignoring both  $\delta^\gamma_{j}\; \delta_{J = j^++j^-}$ yields
  \bea
S(g_1,g_2)&=& \sum_{j^+,j^-,J} d_{j^+} d_{j^-} d_{J}
\int dh\, \chi^{j^+}\bigl(g^+_{1}h (g^+_2)^{-1}\bigr) \crcr
&& \chi^{j^-}\bigl(g^-_{1}h (g^-_2)^{-1}\bigr) \, \chi^{J}(h) \cr
&=& \int dh\, \delta\bigl(g^+_{1}h (g^+_2)^{-1}\bigr)\, \delta\bigl(g^-_{1}h (g^-_2)^{-1}\bigr) \, \delta(h)
\cr&=& \delta\bigl(g^+_{1} (g^+_2)^{-1}\bigr)\, \delta\bigl(g^-_{1}(g^-_2)^{-1}\bigr) \; ,
  \eea
which is the $SO(4)$ BF theory.
\item $\gamma\to\infty$ leads to $j^+=j^-$ and 
\bea
 S(g_1,g_2) =  \sum_j d_{j}^2 \; d_{2j} 
\int dh\, \chi^{j} \bigl( g^+_{1} h(g^+_2)^{-1} \bigr)\, \chi^{j} \bigl(g^-_{1} h(g^-_2)^{-1} \bigr) 
\, \chi^{2j}(h) \; ,
\eea
which is the Barrett Crane spin foam model \cite{Barrett:1997gw}.

\end{itemize}
Returning to eq. (\ref{eq:Sgroup}), note that $S$ admits a single sum representation
\bea
 S(g_1,g_2)&=& \sum_J \Big{[} J(1-\gamma)+1 \Big{]} \Big{[} J(1+\gamma) +1 \Big{]} \Big{[}2J+1 \Big{]}  \cr\cr
  &&\int dh \, \chi^{J\frac{1+\gamma}{2}}\bigl(g^+_{1}h (g^+_2)^{-1}\bigr)\,
\chi^{J\frac{1-\gamma}{2} }\bigl(g^-_{1}h (g^-_2)^{-1}\bigr) \, \chi^{J}(h) \; .
\eea

\section{The EPRL/FK propagator}\label{sec:prop}

In four dimensions the GFT lines have four strands. To build the EPRL/FK propagator one needs to compose
four simplicity projectors, one for each strand, with two gauge invariance projectors, common to all four strands.
The ordinary $SO(4)$ gauge invariance propagator, $T$, corresponding to left invariant fields under 
the diagonal group action on their arguments, i.e. fields satisfying
\bea
\phi(g_1h,g_2h, g_3h, g_4h) = \phi(g_1,g_2,g_3, g_4) \; ,
\eea
has kernel 
\bea
T(\{g_{s}\},\{g'_{s}\} ) =  
\int dh^+ dh^- \prod_{s=1}^4 \delta\bigl(g^+_{s}  h^+ (g'^+_{s})^{-1}\bigr)
\delta\bigl( g^-_{s} h^- (g'^-_{s})^{-1}\bigr) \;,
\eea
where $\{g_s\}$ denotes a collection of four group elements associated to the strands. 
The pair of integration variables $(h^+, h^-)$ is common to all four strands
of a line. The propagator writes
\bea
 C(\{g_s\}; \{g'_s\}) &=& \int \prod_s \bigl( du_s dv_s \bigr) \;  T(\{g_s\}, \{u_s\}) \crcr
                      && \Big{(}\prod_s S(u_s,v_s) \Big{)} \; T(\{v_s\}, \{g'_s\}) \; ,
\eea
or in detail, denoting $\delta_J = \delta_{J=j^++j^-}$ and $h_{\text{in}}^{\pm},h_{\text{out}}^{\pm}$ the
dummy variables corresponding to the two $T$ operators 
\bea
&& C(\{g_s\}; \{g'_s\}) =  \sum_{j_s^+,j_s^-,J_s} d_{j_s^+} d_{j_s^-} d_{J_s} \,\delta^\gamma_{j_s}\; \delta_{J_s}
  \int dh_{\text{in}}^{\pm}  dh_{\text{out}}^{\pm}  \int \prod_s dh_s \crcr
&&\int \prod_s \bigl(du_s^{\pm} dv_s^{\pm} \bigr) \;
\prod_s \delta\bigl(\,g^+_{s}\, h_{\text{in}}^+ \, (u^+_{s})^{-1}\bigr)
\delta\bigl( g^-_{s}\, h_{\text{in}}^- \,(u^-_{s})^{-1}\bigr) \crcr
&&
\prod_s  \chi^{j_s^+}( u_{s}^+\, h_s \,(v_{s}^+)^{-1})\, \chi^{j_s^-}( u_{s}^-\,  h_s \,(v_{s}^-)^{-1}) \, \chi^{J_s}(h_s)
\crcr
&&\prod_s \delta\bigl( v^+_{s} \, h_{\text{out}}^+\, (g'^+_{s})^{-1}\bigr)
\delta\bigl(   v^-_{s} \,h_{\text{out}}^-\, (g'^-_{s})^{-1}\bigr) \; ,
\eea
and integrating over $u_s^{\pm}, v_s^{\pm}$ we get 
\bea
&&C(\{g_s\}; \{g'_s\})=
\sum_{j_s^+,j_s^-,J_s} d_{j_s^+} d_{j_s^-} d_{J_s} \,\delta^\gamma_{j_s}\; \delta_{J_s}
  \int dh_{\text{in}}^{\pm}  dh_{\text{out}}^{\pm}  \int \prod_s dh_s \crcr
&&\prod_s \chi^{j_s^+}(g^+_{s}\, h_{\text{in}}^+\, h_s\, h_{\text{out}}^+ \,  (g'^+_{s})^{-1})\, 
\chi^{j_s^-}(g^-_{s}\, h_{\text{in}}^-\, h_s\,h_{\text{out}}^-\, (g'^-_{s})^{-1})\,
\chi^{J_s}(h_s) \; .
\label{eq:tst}
\eea

\begin{figure}[htb]
\centering{
\includegraphics[width=40mm]{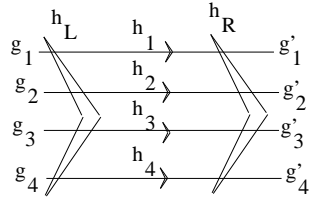}}
\caption{A EPRL/FK line.}
\label{fig:prop}
\end{figure}

A EPRL/FK group field theory line is represented together with all its associated group elements in figure \ref{fig:prop}. 

\section{Feynman amplitudes}\label{sec:ampl}

A Feynman graph of the EPRL/FK group field theory is made of propagators (eq. (\ref{eq:tst})) and vertices
made of trivial conservation $\delta$ functions. Note that the strands are mixed in eq. (\ref{eq:tst})
only through the variables $h^{\pm}_{\text{in},\text{out}}$. In particular, when composing
several such propagators, all the intermediate $g_s$ integrals are factored according to the strands. 
When computing the amplitude of a graph, the integrand is therefore factored into contributions 
associated either to closed strands (called faces) or to open (external) strands.

The composition of two successive strand contributions writes 
\bea
&&\sum_{j_s^+,j_s^-,J_s} d_{j_s^+} d_{j_s^-} d_{J_s} \,\delta^\gamma_{j_s}\; \delta_{J_s}
\sum_{j_{s'}^+,j_{s'}^-,J_{s'}} d_{j_{s'}^+} d_{j_{s'}^-} d_{J_{s'}} \,\delta^\gamma_{j_{s'}}\; \delta_{J_{s'}}
 \int dg'^{\pm}_s \crcr
&&\chi^{j_s^+}(g^+_{s}\, h_{\text{in}}^+\, h_s\, h_{\text{out}}^+ \,  (g'^+_{s})^{-1})\, 
\chi^{j_s^-}(g^-_{s}\, h_{\text{in}}^-\, h_s\,h_{\text{out}}^-\, (g'^-_{s})^{-1})\,
\chi^{J_s}(h_s) \crcr
&&\chi^{j_{s'}^+}(g'^+_{s}\, p_{\text{in}}^+\, p_s\, p_{\text{out}}^+ \,  (g''^+_{s})^{-1})\, 
\chi^{j_{s'}^-}(g'^-_{s}\, p_{\text{in}}^-\, p_s\,p_{\text{out}}^-\, (g''^-_{s})^{-1})\,
\chi^{J_{s'}}(p_s) \; ,
\eea
which, using the orthogonality of characters eq. (\ref{eq:orthchar}) computes to
\bea
&&\sum_{j_s^+,j_s^-,J_s} d_{j_s^+} d_{j_s^-} (d_{J_s})^2 \,\delta^\gamma_{j_s}\; \delta_{J_s} \crcr
&& \chi^{j_s^+}(g^+_{s}\, h_{\text{in}}^+\, h_s\, h_{\text{out}}^+ \,  (p_{\text{in}}^+\, p_s\, p_{\text{out}}^+)
 \,  (g''^+_{s})^{-1}) \crcr
&& \chi^{j_s^-}(g^-_{s}\, h_{\text{in}}^-\, h_s\,h_{\text{out}}^-\, (p_{\text{in}}^-\, p_s\,p_{\text{out}}^-) 
\, (g''^-_{s})^{-1}) \crcr
&&\chi^{J_s}(h_s) \chi^{J_{s}}(p_s) \; .
\eea

In an arbitrary Feynman amplitude we therefore have one surviving independent sum per face of the graph
and one per external strand. 

To write the full amplitude of a graph $\mathcal{G}$ we introduce some notations. We denote the two couples of 
in and out variables of a line $l$ by $h^{\pm}_{\text{in};l}$ and $h^{\pm}_{\text{out};l}$.
We denote $\partial f$ the set of lines of the boundary of the face $f$ and $|\partial f|$ its cardinal. 
For each line $l\in \partial f$ we have a variable $h_{lf}$ (corresponding to $h_s$ 
in eq. (\ref{eq:tst})). 
Furthermore, we denote $\epsilon_{lf} $ the incidence matrix of lines within faces \cite{KMRTV,Geloun:2010nw}, which is
$0$ if $l\notin\partial f$ and $1$ (or $-1$) if $l \in \partial f$ and the orientations of $l$ and $f$ 
coincide (or not). Finally, denoting ${\mathcal L}_{{\mathcal G}}$ the set of lines and ${\mathcal F}_{{\mathcal G}}$ 
the set of faces of $\mathcal{G}$, the amplitude writes
\bea\label{eq:ampli}
&& A_{\mathcal G} (\{g_s^+\}, \{g_s^-\})=
\sum_{j_f^+,j_f^-,J_{lf} }  
\Big{(}\prod_{f\in {\mathcal F}_{{\mathcal G}} } d_{j_f^+} d_{j_f^-} \, \delta^\gamma_{j_f} 
 \; \big{(} \prod_{l\in\partial f} d_{J_{lf}} \delta_{J_{lf}= j_f^++j_f^-}  \big{)} \Big{)} \\
&&
\int \Bigl{[} \prod_{l\in {\mathcal L}_{{\mathcal G}}}\; dh_{\text{in};\,l}^{\pm}  dh_{\text{out};\,l}^{\pm} \Big{]}
\int \Bigl{[}\prod_{\stackrel{l\in {\mathcal L}_{{\mathcal G}},f\in {\mathcal F}_{{\mathcal G}}}{l\in \partial f}}
 dh_{lf}\,\Bigr{]} \;  
\Big{[} \prod_{\stackrel{l\in {\mathcal L}_{{\mathcal G}},f\in {\mathcal F}_{{\mathcal G}}}{l\in \partial f}}  
\chi^{J_{lf}} (h_{lf}) \Big{]} \crcr
&& \prod_{f\in {\mathcal F}_{{\mathcal G}} } \Big{[} \chi^{j^+_f} \Big{(} \prod_{l \in \partial f}\,
(h_{\text{in};\,l}^+\,h_{lf} \,  h_{\text{out};\,l}^+)^{\epsilon_{lf}} \Big{)}  \; \chi^{j^-_f} \Big{(} \prod_{l \in \partial f} \, 
(h_{\text{in};\,l}^-\,h_{lf} \,  h_{\text{out};\,l}^-)^{\epsilon_{lf}} \Big{)} \Big{]}   \; ,\nonumber
\eea
where for external, open faces, with group elements at the endpoints 
$g_s^{\pm}$ and $g_s'^{\pm}$, the last line is replace by 
\bea\label{eq:ampliex}
&& \chi^{j^+_f} \Big{[}(g_s^+)^{\epsilon_{ef}} \Big{(} \prod_{l \in \partial f}\,
(h_{\text{in};\,l}^+\,h_{lf} \,  h_{\text{out};\,l}^+)^{\epsilon_{lf}} \Big{)} (g_s'^+)^{\epsilon_{ef}}\Big{]}  
\crcr
&& \chi^{j^-_f} \Big{[} (g_s^-)^{\epsilon_{ef}} \Big{(} \prod_{l \in \partial f} \, 
(h_{\text{in};\,l}^-\,h_{lf} \,  h_{\text{out};\,l}^-)^{\epsilon_{lf}} \Big{)} (g_s'^-)^{\epsilon_{ef}}\Big{]}   \; ,
\eea
with $\epsilon_{ef}$ the incidence matrix of external points with faces.

Before concluding this section lest us note that one could use an arbitrary power of the $TST$ operator as propagator, 
since any power would effectively implement the Plebanski constraints.  
We can presumably, in this category of theories generalizing EPRL/FK, always
adjust the power $k$ so as to find a just renormalizable theory\footnote{We acknowledge D. Oriti for making this remark
during a most enjoyable conference in AEI, Golm.}. The amplitudes of such a model will have the same form as
eq. (\ref{eq:ampli}), but with extra insertions of intermediate variables
 $h^{\pm}_{\text{in},\text{out}}$ and $h_{lf}$ along the faces improving the power counting of the theory.

\section{Subtraction, locality, and all that}\label{sec:subtr}

Starting form the eq. (\ref{eq:ampli}) of the Feynman amplitude of a graph one can address the subtraction 
of divergences in this theory. This procedure is straightforward with 
the definition of locality proposed in \cite{KMRTV}.

Take the example of a two point function. The amplitude of a connected graph writes in terms of the 
amplitude of the amputated graph as
\bea
 A_{\cG}(\phi) = \int dg_s dg_s' \; \phi(\{g_s\}) \; \phi( \{g_s' \}) \; A_{\cG} ({\{g_s\}, \{g_s'\}} ) \; ,
\eea
the leading (``mass'') divergence is immediately identified by Taylor developing ``at zeroth order'' 
the field $\phi( \{g_s' \})$\footnote{As always, sub leading divergences are more difficult to 
extract (one needs to push further the Taylor development of the external fields), and is deferred 
for further work.} around $\{g'_s\}=\{g_s\}$ 
\bea
 A_{\cG}(\phi) &=& \int dg_s \; \phi(\{g_s\}) \; \phi(\{g_s\}) \int dg_s' \; A_{\cG} ({\{g_s\}, \{g_s'\}} ) 
 +\text{ sub leading }
\crcr
&&=\delta \mu_{\cG} \int dg_s \; \phi(\{g_s\}) \; \phi(\{g_s\}) + \text{sub leading}
\; .
\eea

Taking into account eq. (\ref{eq:ampliex}), we note that the integration over the external field $g_s'^{\pm}$ 
fixes all $j^+_{f}$, $j^-_{f}$ and $J_{lf}$ to $0$ and the external strand contribution drops out of eq. 
(\ref{eq:ampli}). In general, the leading divergence of any graph $\cG$ is therefore obtained by 
integrating eq. (\ref{eq:ampli}) ignoring the external strands.

\begin{figure}[htb]
\centering{
\includegraphics[width=40mm]{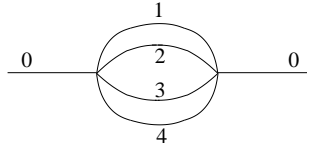}}
\caption{A graph exhibiting a mass divergence.}
\label{fig:self}
\end{figure}

Take the example of the graph $\cG$ drawn schematically in figure \ref{fig:self}. All lines
have parallel strands, and are oriented from left to right. We denote the lines $1$ to $4$
(which can be interpreted as colors in a colored model), and the face by the couple of labels
of the lines composing them. The set of internal faces of this graph is therefore 
$f=\{ f_{12},f_{13},f_{14},f_{23},f_{24},f_{34}\}$.

The mass divergence of $\cG$ writes
\bea\label{eq:monster}
\delta\mu_{\cG} &=& \sum_{j_{12}^+,j^-_{12}, J_{1,12},J_{2,12} ,\dots } 
\Big{(} d_{j^+_{12}}d_{j^-_{12}}  \delta^{\gamma}_{j_{12}}  \big{(} d_{J_{1,12}} \delta_{J_{1,12}} 
d_{J_{2,12}} \delta_{J_{2,12}} \big{)} \Big{)}\dots 
\crcr
&&  \int dh^{\pm}_{\text{in},1} dh^{\pm}_{\text{out},1} \dots \int dh_{1,12} dh_{2,12} \dots \chi^{J_{1,12}}(h_{1,12})
 \chi^{J_{2,12}}(h_{2,12}) \dots 
\crcr
&&
 \chi^{j^+_{12}}(h_{\text{in},1}^+ h_{1,12} h_{\text{out},1}^+ (h_{\text{in},2}^+ h_{2,12} h_{\text{out},2}^+)^{-1} ) \crcr
&&\chi^{j^-_{12}}(h_{\text{in},1}^- h_{1,12} h_{\text{out},1}^- (h_{\text{in},2}^- h_{2,12} h_{\text{out},2}^-)^{-1} ) \dots \; .
\eea
In eq. (\ref{eq:monster}) we have 6 independent sums, 16 integrals over line $h^{\pm}_{\text{in},\text{out}}$ variables,
12 integrations over $h_{i,ij}$ strand variables of a product of 24 characters. Unsurprisingly, the full 
evaluation of the amplitude of this graph is somewhat involved (see \cite{Perini:2008pd,KMRTV}), but 
fortunately one can derive its degree of divergence in our group representation relatively straight 
forward.

Divergences arise for large values of the spin labels $j^{\pm}, J$, thus we cutoff all the sums by some 
sharp cutoff $\Lambda$. Each $d_j^{\pm},d_J$ factor in the first line of eq. (\ref{eq:monster}) will bring 
a factor $\Lambda$. The integrals over the characters are of the form 
\bea \label{eq:saddle}
 \int \prod_{j=1}^n d\theta_j \Big{(} \sin\frac{\theta_j}{2} \Big{)}^2  \int_{S^2} d q_j \; F(\Lambda,\theta_j, q_j) 
\; ,
\eea
where we have used the representation eq. (\ref{eq:haar}) of the Haar measure over $SU(2)$. The integrals over the normals 
$q_j$ are bounded by 1 and will be ignored. The integrals over $\theta_j$ will be evaluated by 
some saddle point approximation. The saddle point equations are $\theta_j=\theta^s_j$ with
\bea
  \theta_p^s = 0 \quad \forall p\le k \; , \qquad  \theta_p^s \neq 0 \quad \forall p>k \; .
\eea
The behavior of eq. (\ref{eq:saddle}) is strongly dependent of $k$. In fact, when translating at the saddle point
$x_j = \theta_j - \theta_j^s$, eq. (\ref{eq:saddle}) writes
\bea
 \int \prod_{j=1}^k \Big{(}\sin\frac{x_j}{2}\Big{)}^2 dx_j 
\prod_{j=k+1}^n \Big{(}\sin\frac{x_j+ \theta^s_j}{2}\Big{)}^2 d x_j  F(\Lambda, x+\theta^s) \; ,
\eea
and performing the rescaling $ x_j = \frac{u_j}{\sqrt{\Lambda}}$ close to the saddle point we get
\bea\label{eq:scaling}
&& \int \prod_{j=1}^k \Big{(}\sin\frac{u_j}{2\sqrt{\Lambda}}\Big{)}^2 \frac{du_j}{\sqrt{\Lambda}} 
\prod_{j=k+1}^n \Big{(}\sin\frac{\frac{u_j}{\sqrt{\Lambda}}+ \theta^s_j}{2}\Big{)}^2 
\frac{d u_j}{\sqrt{\Lambda}}  F(\Lambda, \frac{u}{\sqrt{\Lambda}}+\theta^s) \crcr
&&\approx  \frac{1}{(\sqrt{\Lambda})^{3k}} \frac{1}{(\sqrt{\Lambda})^{n-k} } \crcr 
&&\int \prod_{j=1}^k \frac{u_j^2}{4} du_j 
\prod_{j=k+1}^n \Big{(}\sin\frac{\theta^s_j}{2}\Big{)}^2 
d u_j \; F(\Lambda, \frac{u}{\sqrt{\Lambda}}+\theta^s) \; ,
\eea
and the remaining integral gives no extra scaling in $\Lambda$. Therefore the scaling of
eq. (\ref{eq:saddle}) is fixed by $n$ (the number of integration variables) and $k$
(the number of directions with saddle point equation $\theta_j=0$). 

For the graph of figure \ref{fig:self}, we change variables to
\bea
&&(\tilde h^{+}_{in;2})^{-1} = h_{in,1}^+ h_{1,12} h_{out,1}^+ (h_{out,2}^+)^{-1} h_{2,12}^{-1} (h_{in,2}^+)^{-1}  \crcr
&&(\tilde h^{+}_{in;3})^{-1} = h_{in,1}^+ h_{1,13} h_{out,1}^+ (h_{out,3}^+)^{-1} h_{3,13}^{-1} (h_{in,3}^+)^{-1}  \crcr
&&(\tilde h^{+}_{in;4})^{-1} = h_{in,1}^+ h_{1,14} h_{out,1}^+ (h_{out,4}^+)^{-1} h_{4,14}^{-1} (h_{in,4}^+)^{-1}  \;,
\eea
and similarly for the $-$ variables. This brings the contribution of the faces $f_{12},f_{13},f_{14}$ into the form 
\bea
&&\chi^{j^+_{12}} \bigl((\tilde h^{+}_{in;2})^{-1} \bigr) \chi^{j^+_{13}}\bigl((\tilde h^{+}_{in;3})^{-1} \bigr)
\chi^{j^+_{14}}\bigl((\tilde h^{+}_{in;4})^{-1} \bigr) \crcr
&&\chi^{j^-_{12}} \bigl((\tilde h^{-}_{in;2})^{-1} \bigr) \chi^{j^-_{13}}\bigl((\tilde h^{-}_{in;3})^{-1} \bigr)
\chi^{j^-_{14}}\bigl((\tilde h^{-}_{in;4})^{-1} \bigr) \; ,
\eea 
while the ($+$ part) contribution of the face $f_{23}$ becomes 
\bea
&&\tilde h^{+}_{in;2} \; h_{in,1}^+ h_{1,12} h_{out,1}^+ (h_{out,2}^+)^{-1} h_{2,12}^{-1}
\crcr
&& h_{2,23} h_{out,2}^+ (h_{out,3}^+)^{-1} h_{3,23}^{-1} 
\crcr 
&&  h_{3,13} h_{out,3}^+ (h_{out,1}^+)^{-1}   h_{1,13}^{-1} (h_{in,1}^+)^{-1}
(\tilde h^{+}_{in;3})^{-1} \; ,
\eea
and similarly for the faces $f_{24}$, $f_{24}$ and $f_{34}$. Note that all the remaining variables,
($h^+_{\text{in};1}$ and  $h^{+}_{\text{out}; 1}$ , $h^{+}_{\text{out}; 2}$, $h^{+}_{\text{out}; 3}$, $h^{+}_{\text{out}; 4}$) 
appear always in pairs $h$, $h^{-1}$. 

The integration variables $h_{lf}$ and $\tilde h$ appear explicitly as arguments of some character  
\bea
 \int dh \; \chi^{j}(h) F(h,\dots) \; .
\eea
For all this variables, and the associated $\theta_h^s\neq 0$ as
\bea
 \int dh \; \chi^{j}(h) F(h)  = \int d\theta_h \; \sin\frac{\theta_h}{2} \; \sin\frac{(2j+1)\theta_h}{2} \; 
F(\theta_h,\dots) \; ,
\eea
and the integrand is exactly zero at $\theta_h=0$. It is easy to check that the remaining group elements, as they appear only
in pairs $h$, $h^{-1}$ have $\theta_h=0$ at the saddle.
We therefore have $12\times h_{lf}+3\times \tilde h^+ + 3 \times \tilde h^-$ variables with $\theta^s\neq 0$
and $5\times h^+ + 5 \times h^-$ variables with $\theta^s=0$. The scaling at the saddle point is, according to eq. (\ref{eq:scaling}),
\bea
  \frac{1}{\sqrt{\Lambda}^{3\times 10}} \frac{1}{\sqrt{\Lambda}^{18}} = \Lambda^{-24} \; .
\eea 
In eq. (\ref{eq:monster}) we count 6 independent sums and 24 factor $d_{j^+}$, $d_{j^-}$ and $d_{J}$, hence 
\bea
 \delta \mu \approx \sum_{6 \times} \Lambda^{24} \Lambda^{-24} \approx \Lambda^6 \; ,
\eea
which coincides with the results of \cite{Perini:2008pd,KMRTV}.

This power counting argument can be used to also derive for instance the degree of divergence
of the graph $\cG$ for the BF model with $SU(2)$ group ($\gamma=1$). In this case the $-$ variables
are absent and we have $n=20$, $k=5$ and we recover the well known scaling
\bea
 \sum_{6 \times } \Lambda^{18} \frac{1}{\sqrt{\Lambda}^{3 \times 5}} \frac{1}{\sqrt{\Lambda}^{15}}=\Lambda^9 \; .
\eea

For an arbitrary graph the saddle point analysis becomes more involved, and the scaling is
influenced both by the position of the saddle point in the $\theta$ space and by the presence of 
degenerate directions. A precise analysis is in progress \cite{BGR}.

As a final observation, note that the Barett Crane model $\gamma \to \infty$ has exactly the same 
divergences as the EPRL/FK model. We expect however that the sub leading divergences can 
be subtracted (in some ``wave function'' renormalization) only for the EPRL/FK models, leading to 
a non trivial flow of the Immirzi parameter.

\section*{Acknowledgements}

Research at Perimeter Institute is supported by the Government of Canada through Industry 
Canada and by the Province of Ontario through the Mi-nistry of Research and Innovation.

\section*{Appendix}

\appendix

\section{$SU(2)$ coherent states }\label{app:A}
\renewcommand{\theequation}{A.\arabic{equation}}
\setcounter{equation}{0}

An element $g$ of $SU(2)$ writes $ g=e^{i \frac{\theta}{2} \vec{k} \cdot \vec{\sigma}} $
where $\vec{\sigma}=(\sigma_{x},\sigma_{y},\sigma_{z})$ are the
Pauli matrices. In this parametrization the Haar measure on $SU(2)$
is 
\bea \label{eq:haar}
\int_{SU(2)}\, d\mu(g)  = \frac{1}{2\pi} 
\int_{0}^{4\pi} d\theta\sin^2 \frac{\theta}{2}\int_{S^2} d k  \;.
\eea
Alternatively, elements of $SU(2)$ can  be parametrized by Euler angles (in $z-y-z$ order)
\bea
g= e^{-i\alpha \sigma_z}\,e^{-i\beta \sigma_y} e^{-i\gamma \sigma_z} \; ,
\eea
representing the rotation of angle $\gamma$ around the direction 
\bea
 \vec n = (\sin\beta \, \cos\alpha,\;  \sin\beta\, \sin\alpha, \; \cos\beta) \;,
\eea
and the Haar measure writes in terms of Euler angles as
\bea
\int_{SU(2)}\,  d\mu(g) = \frac{1}{2\pi} \int_{0}^{2\pi}d\gamma \; \int_{S^2}  dn \; ,
\eea
where we use the normalized measure on the sphere $S^2$ 
\bea
\int_{S^2} dn =  \frac{1}{4\pi} \int_{0}^{\pi} d\beta \sin\beta \int_{0}^{2\pi}d\alpha.
\eea

In the spin $j$ representation space of $SU(2)$, $H_j= \{|j,m\rangle,\; |m|\leq j\}$,
the Wigner matrix representing $g$ writes in Euler angles 
\bea \label{eq:Wigner}
D^{j}_{pq}(g) = \langle j,p \vert g^j\vert j,q\rangle =
e^{-i\alpha p} d^{j}_{pq}(\beta)  e^{-i\gamma q} \; .
\eea
The coherent states on $SU(2)$ \cite{perelomov} are indexed by a vector $\vec n$ 
\bea\label{eq:coh}
 \vert j, n \rangle \equiv \sum_{p} D^j_{pj}(\alpha,\beta,0) | j,p \rangle \; .
\eea
Note that in the definition of the coherent states one uses Wigner matrices with 
$\gamma$, the third Euler angle, set to zero. When dealing with coherent states one
needs to reestablish the dependence of this Euler angles and transform integrals over
the vector $\vec n$ into integrals over the $SU(2)$ group. Consider for instance the
integral 
\bea\label{eq:id}
 d_j \int dn \, | j,n \rangle \langle j,n |
& =&  \frac{d_j}{4\pi} \int_{0}^{2\pi} d\alpha \int_{0}^{\pi} \sin \beta d\beta \nonumber \\
&&\sum_{p,s} D^j_{pj}(\alpha,\beta,0) | j,p \rangle
\overline{D^j_{sj}(\alpha,\beta,0)} \langle j,s | \; .
\eea
We first add an extra normalized integral over a fictitious variable, $\chi$, 
\bea
&& \frac{d_j}{4\pi} 
\int_{0}^{2\pi} d\alpha \int_{0}^{\pi} \sin \beta d\beta\;
\frac{1}{2\pi} \int_0^{2\pi} d\chi  \nonumber \\
&&\sum_{p,s} D^j_{pj}(\phi,\psi,0) | j,p \rangle e^{-i \chi j} e^{i\chi j}
\overline{D^j_{sj}(\phi,\psi,0)} \langle j,s | .
\eea
But, due to eq. (\ref{eq:Wigner}),
$ D^j_{pj}(\phi,\psi,0) e^{-i \chi j} = D^j_{pj}(\phi,\psi,\chi)$. 
Moreover, the integrals over $\alpha, \beta$ and $\chi$ reproduce the Haar measure on $SU(2)$,
hence eq. (\ref{eq:id}) becomes
\bea
d_j \int dg \sum_{p,s} D^j_{pj}(\phi,\psi,\chi) \overline{D^j_{sj}(\phi,\psi,\chi)}
| j,p \rangle \langle j,s |,
\eea
and using the orthogonality of the Wigner matrices 
\bea
 \int dg \; D^j_{pj}(g) \overline{D^j_{sj}(g)} = \frac{1}{d_j} \, \delta_{p,s} \; 
\eea
one concludes that the coherent states yield a resolution of the identity
\bea
 d_j \int dn \, | j,n \rangle \langle j,n | = \sum_p | j,p \rangle \langle j,p | = \mathbb{I}_j \; .
\eea

\section{Evaluation of ${\mathcal I}(\vec  j , \vec m_1,\vec m_2)$}\label{app:B}

\renewcommand{\theequation}{B.\arabic{equation}}
\setcounter{equation}{0}

In this appendix we compute the integral ${\mathcal I}(\vec  j , \vec m_1,\vec m_2)$ of
eq. (\ref{eq:Idef}). 
\bea
{\mathcal I}(\vec  j , \vec m_1,\vec m_2) &&=
\int dn dk \; \langle  \vec  j , \vec m_1 \vert \Big{(} \vert j^+,  n \rangle \otimes \vert j^-, n \rangle\Big{)}
\crcr && 
\langle   j^+ + j^- ,  n \vert     j^+ + j^-  , k \rangle 
\Big{(}\langle  j^+, k \vert \otimes \langle j^-, k\vert \Big{)} \vert \vec j, \vec m_2  \rangle \;.
\label{eq:ijmn}
\eea
Using the definition of coherent states eq. (\ref{eq:coh}) and inserting judiciously phases
in the new fictitious variables (generalizing straightforward the manipulation in appendix \ref{app:A})
eq. (\ref{eq:ijmn}) writes 
\bea
{\mathcal I}(\vec  j , \vec m_1,\vec m_2)
=&&
\sum_{r} \int dg dg'\, D^{j^+}_{m_1^+j^+}(g)D^{j^-}_{m^-_1j^-}(g)
\overline{D^{j^+ + j^-}_{r(j^+ + j^-)}(g)} \crcr
&& D^{j^+ + j^-}_{r(j^+ + j^-)}(g')
\overline{D^{j^+}_{m^+_2 j^+ }(g')} \; \overline{D^{j^-}_{m^-_2 j^- }(g')} \; .
\eea
Under hermitian and complex conjugation the Wigner matrices satisfy the relations
$D^j_{mn}(g^{-1})= \overline{ D^{j}_{nm}(g)}= (-1)^{n-m} D^{j}_{-n-m}(g) $,
thus
\bea
{\mathcal I}(\vec  j , \vec m_1,\vec m_2)
&=&
\sum_{r} (-1)^{r-j^+-j^-+m^+_2- j^++m^-_2- j^-} \crcr
&& \int dg  D^{j^+}_{m_1^+j^+}(g)D^{j^-}_{m^-_1j^-}(g)
 D^{j^+ + j^-}_{-r-(j^+ + j^-)}(g)\cr \crcr
&& \int dg' D^{j^+ + j^-}_{r(j^+ + j^-)}(g')
D^{j^+}_{- m^+_2 -j^+ }(g')  D^{j^-}_{-m^-_2 -j^- }(g') \; .
\eea
The group integrals of products of three Wigner matrices compute in terms
of Wigner 3j symbols \cite{messiah}
\bea
&&\int dg\; D^{j_1}_{m_{1}n_{1}}(g)\, D^{j_2}_{m_{2}n_{2}}(g)\, D^{J}_{MM'}(g) \crcr
&&  = \left(\begin{array}{ccc}
  j_1  & j_2 & J\\
  m_1 & m_2 & M
\end{array}\right)
\left(\begin{array}{ccc}
  j_1  & j_2 & J\\
  n_1 & n_2 & M'
\end{array}\right) \; ,
\eea
thus 
\bea
{\mathcal I}(\vec  j , \vec m_1,\vec m_2)
&=&(-1)^{m^+_2+m^-_2 - 2(j^++ j^-)} \sum_{r} (-1)^{r}
\left(\begin{array}{ccc}
  j^+ & j^- & j^+ +j^-\\
  m^+_1 & m^-_1 & -r
\end{array}\right)\crcr
&&
\left(\begin{array}{ccc}
  j^+  & j^- & j^+ +j^-\\
   j^+ & j^-  & -(j^+ +j^-)
\end{array}\right)
\left(\begin{array}{ccc}
  j^+  & j^- & j^+ +j^-\\
  -m^+_2 & -m^-_2 & r
\end{array}\right)\crcr
&&
\left(\begin{array}{ccc}
  j^+  & j^- & j^+ +j^- \\
  -j^+ & -j^- & j^+ +j^-
\end{array}\right) \;,
\label{eq:mat3j}
\eea
which writes using the symmetry properties of the $3j$ symbols 
\bea
{\mathcal I}(\vec  j , \vec m_1,\vec m_2) &=& 
\left(\begin{array}{ccc}
  j^+  & j^- & j^+ +j^-\\
   j^+ & j^-  & -(j^+ +j^-)
\end{array}\right)^2  
\sum_r  (-)^{r+m^+_2+m^-_2 } \\
&&
  \left(\begin{array}{ccc}
  j^+ & j^- & j^+ +j^-\\
  m^+_1 & m^-_1 & -r
\end{array}\right) \left(\begin{array}{ccc}
  j^+  & j^- & j^+ +j^-\\
   -m^+_2 & -m^-_2 & r
\end{array}\right) \; .\nonumber
\eea
Taking into account the evaluation of particular $3j$ symbols
\bea
\left(\begin{array}{ccc}
  j^+  & j^- & j^+ +j^- \\
  j^+ & j^- & -(j^+ +j^-)
\end{array}\right)  =\frac{(-1)^{2j^+}}{\sqrt{2(j^+ +j^-)+1}} \; ,
\eea
we get 
\bea
 {\mathcal I}(\vec  j , \vec m_1,\vec m_2) = \frac{1}{d_{j^++j^-}} && \sum_r  (-)^{r+m^+_2+m^-_2 } 
  \left(\begin{array}{ccc}
  j^+ & j^- & j^+ +j^-\\
  m^+_1 & m^-_1 & -r
\end{array}\right) \crcr
&&
\left(\begin{array}{ccc}
  j^+  & j^- & j^+ +j^-\\
   -m^+_2 & -m^-_2 & r
\end{array}\right) \; .
\eea

Also note that, according to \cite{messiah},
\bea
&&\left(\begin{array}{ccc}
  j^+  & j^- & j^+ +j^-\\
   -m^+_2 & -m^-_2 & r
\end{array}\right)  =  \frac{(-1)^{j^+-j^- -r }}{\sqrt{d_{j^++j^-}}}
\sqrt{\frac{(2j^+)!(2j^-)!}{(2j^+ +2 j^-)!}}\crcr
&&
\sqrt{\frac{(j^+ +j^- + r)!\, (j^+ +j^- - r)! }{(j^+ + m^+_2)!(j^+-m^+_2)!(j^- + m_2^-)!(j^--m_2^- )!}} \crcr
&&= (-1)^{-2r} \left(\begin{array}{ccc}
  j^+  & j^- & j^+ +j^-\\
   m^+_2 & m^-_2 &  -r
\end{array}\right) \; .
\eea
By the selection rules, the $3j$ symbol is zero unless $ m^+_2 + m^-_2 -r =0$, hence we finally get
\bea
 {\mathcal I}(\vec  j , \vec m_1,\vec m_2) = \frac{1}{d_{j^++j^-}} && \sum_r 
  \left(\begin{array}{ccc}
  j^+ & j^- & j^+ +j^-\\
  m^+_1 & m^-_1 & -r
\end{array}\right) \crcr
&&
\left(\begin{array}{ccc}
  j^+  & j^- & j^+ +j^-\\
  m^+_2 & m^-_2 & -r
\end{array}\right) \; ,
\eea
which can be rewritten as
\bea
{\mathcal I}(\vec j, \vec m_1, \vec m_2) = \frac{1}{d_{j^+ +j^-}}
\sum_{r}\int dh \; D^{j^+}_{m^+_1,m^+_2} (h) 
D^{j^-}_{m^-_1,m^-_2} (h)D^{j^++j^-}_{-r,-r} (h) \; .
\label{ijmn2}
\eea

\end{document}